# Ultrahigh efficient spin-orbit torque magnetization switching in all-sputtered topological insulator - ferromagnet multilayers

Tuo Fan[1], Nguyen Huynh Duy Khang[1,2], Soichiro Nakano[1], Pham Nam Hai[1,3,4]

[1]Department of Electrical and Electronic Engineering, Tokyo Institute of Technology,

2-12-1 Ookayama, Meguro, Tokyo 152-8550, Japan

[2]Department of Physics, Ho Chi Minh City University of Education, 280 An Duong Vuong Street,

District 5, Ho Chi Minh City 738242, Vietnam

[3]Center for Spintronics Research Network (CSRN), The University of Tokyo,

7-3-1 Hongo, Bunkyo, Tokyo 113-8656, Japan

[4]CREST, Japan Science and Technology Agency,

4-1-8 Honcho, Kawaguchi, Saitama 332-0012, Japan

**Abstract**

Spin-orbit torque (SOT) magnetization switching of ferromagnets with large perpendicular magnetic anisotropy has a great potential for the next-generation non-volatile magnetoresistive random-access memory (MRAM). It requires a high-performance pure spin current source with a large spin Hall angle and high electrical conductivity, which can be fabricated by a mass production technique. In this work, we demonstrate ultrahigh efficient and robust SOT magnetization switching in all-sputtered BiSb topological insulator – perpendicularly magnetized Co/Pt multilayers. Despite fabricated by the industry-friendly magnetron sputtering instead of the laboratory molecular beam epitaxy, the topological insulator layer, BiSb, shows a large spin Hall angle of $\theta_{\text{SH}}$ = 12.3 and high electrical conductivity of $\sigma = 1.5 \times 10^5$ $\Omega^{-1}$ m$^{-1}$. Our results demonstrate the mass production capability of BiSb topological insulator for implementation of ultralow power SOT-MRAM and other SOT-based spintronic devices.

Embedded non-volatile memories have great impacts on energy-efficient electronics, including Internet-of-Thing, Artificially Intelligent (AI), among others. To be successful, non-volatile memories have to satisfy several requirements, such as high writing endurance, high capacity, high speed, and low fabrication cost. Among many emerging non-volatile memory technologies, magnetoresistive random-access memory (MRAM) is one of the most promising that have gained development commitment from several leading semiconductor companies. The latest MRAM technology using the sophisticated spin-transfer-torque (STT) writing technique has just been commercially available very recently, but already found various important applications, such as highly efficient AI chips. However, in STT-MRAM, a large writing current has to be injected directly to magnetic tunneling junctions (MTJs), which leads to reliability issues such as accelerated aging of the oxide tunnel barrier.[1] In addition, large writing currents require large driving transistors, making it difficult to increase the bit density of STT-MRAM beyond the 1 Gbit capacity. Recently, the spin-orbit-torque (SOT) technique has emerged as a promising writing method for the next generation MRAM.[2] In SOT-MRAM, a charge current flowing in a non-magnetic layer with large spin-orbit interaction can generate a pure spin current by the spin Hall effect. The pure spin current is then injected to the magnetic free layer for magnetization switching. The relationships between the spin current $I_S$ and the charge current $I_C$ is given by $I_S = (\hbar/2e)(L/t)\theta_{SH}I_C$, where $L$ is the MTJ size, and $t$ the thickness of the spin Hall layer, and $\theta_{SH}$ is the spin Hall angle. Theoretically, the charge-to-spin conversion efficiency $(L/t)\theta_{SH}$ in SOT-MRAM can be larger than unity, meaning that lower driving currents can be expected. Furthermore, since there is no current flowing into the MTJs, reliability can be significantly improved. Finally, since the spin-polarization of the pure spin current is perpendicular to the magnetization direction of the free magnetic layer, the spin torque is maximized and the magnetization can switch very fast (< ns) in SOT-MRAM with perpendicular magnetic anisotropy (PMA).[3] Because of those merits of SOT-MRAM comparing with STT-MRAM, there have been huge efforts to find spin Hall materials with large $\theta_{SH}$ and high electrical conductivity for SOT-MRAM implementation. Heavy metals, such as Pt,[4,5] Ta,[2] and W,[6] have been studied extensively as candidates for the spin Hall layer in SOT-MRAM, since they have been used in STT-MRAM manufacturing. However, $\theta_{SH}$ of heavy metals is of the order of ~ 0.1, and the typical critical switching current density in heavy metals/ferromagnet bilayers with PMA is of the order of $10^7$ Acm$^{-2}$ to $10^8$ Acm$^{-2}$, which is too high for Si electronics.[7] Thus, finding a material with large $\theta_{SH}$ of the order of 10 is of emergent need.

Recently, very large $\theta_{SH}$ (>1) has been observed in topological insulators (TIs), making them very promising for SOT-MRAM.[8,9] TIs are quantum materials having gaped bulk states but Dirac-like metallic surface states[10,11,12] with spin-momentum locking,[13,14] owning to their large spin-orbit interaction and band structure topology. The large $\theta_{SH}$ of TIs has been clearly shown to originate from the surface states rather than the bulk states.[15] However, in most TI thin films, the limited surface density of states restricts the electrical conductivity $\sigma$ to the order of ~ $10^4$ $\Omega^{-1}$ m$^{-1}$ (for example, $\sigma$ ~ 5.7×$10^4$ $\Omega^{-1}$m$^{-1}$ for $Bi_2Se_3$ and 1.8×$10^4$ $\Omega^{-1}$m$^{-1}$ for $(Bi_{0.07}Sb_{0.93})_2Te_3$), which results in a large shunting current when in contact with other metallic layers. Second, most TI thin films studied so far are single crystalline films grown epitaxially by the laboratory molecular beam epitaxy (MBE) technique on dedicated III-V semiconductor substrates, which is not compatible with mass production. Furthermore, ultralow power SOT magnetization switching at room-temperature with current densities of the order of $10^5$ – $10^6$ Acm$^{-2}$ have been demonstrated mostly in TI/magnetic layers whose magnetic anisotropy is very small, such as in a $Bi_2Se_3$/NiFe bilayer with nearly zero in-plane magnetic anisotropy,[16] or in a $Bi_2Se_3$/CoTb bilayer with a small magnetization of 200 emu/cc and a PMA field of less than 2 kOe.[17] For TI being a practical material for SOT-MRAM, the following three minimum requirements must be satisfied: (1) a large spin Hall angle of the order of 10, (2) large electrical conductivity $\sigma$ of order of $10^5$ $\Omega^{-1}$ m$^{-1}$, and (3) can be deposited using industry-friendly techniques, such as sputtering deposition. The closest attempt is $Bi_xSe_{1-x}$ thin films deposited by the sputtering technique, which shows a promisingly large $\theta_{SH}$ = 8.7-18.6 but with the expense of reduced electrical conductivity ($\sigma$ ~ 7.8×$10^3$ $\Omega^{-1}$m$^{-1}$).[18] Thus, the three requirements for a practical spin Hall material for SOT-MRAM have not yet been achieved.

Among various TI material candidates, $Bi_{1-x}Sb_x$ (0.07 < $x$ < 0.22) is the best hope. BiSb is the first discovered three-dimensional TI with small bulk bandgap (~ 20 meV) and high bulk conductivity $\sigma$ of 4–6.4×$10^5$ $\Omega^{-1}$ $m^{-1}$.[13,14,19] In thin films, the quantum size effect significantly increases the band gap of BiSb, so that the current flows mostly on the surface when the thickness reaches 10 nm.[20] Thanks to the multi-surface states, $\sigma$ of BiSb thin films is as high as 2.5×$10^5$ $\Omega^{-1}$ $m^{-1}$. Furthermore, a giant spin Hall effect with $\theta_{SH}$ ~ 52 has been observed in BiSb(012) thin films in junctions with MnGa grown on GaAs substrates by MBE.[21] Nevertheless, following works on sputtered polycrystalline BiSb yields a maximum $\theta_{SH}$ of only 1.2 (ref. 22), or even no spin Hall effect.[23] Therefore, it is very important to demonstrate the three requirements for sputtered BiSb for any realistic applications to SOT-based spintronic devices.

In this work, we demonstrate ultrahigh efficient SOT magnetization switching in all-sputtered BiSb – (Co/Pt) multilayers with large PMA. We show that the sputtered BiSb has a large spin Hall angle of $\theta_{SH}$ = 12.3 and high electrical conductivity of $\sigma$ = 1.5 × $10^5$ $\Omega^{-1}m^{-1}$, thus satisfying all the three requirements for SOT-MRAM implementation. Despite the large PMA field of 5.2 kOe of the (Co/Pt) multilayers, we achieve robust SOT magnetization at a low current density of 1.5×$10^6$ $Acm^{-2}$. Our results demonstrate the potential of BiSb topological insulator for mass production of ultralow power SOT-MRAM and other SOT-based spintronic devices.

**High quality topological insulator - perpendicularly magnetized ferromagnetic multilayers**

The first step in this study is to demonstrate that it is possible to use the sputtering deposition technique for fabrication of high-quality topological insulator - perpendicularly magnetized ferromagnetic multilayers. Figure 1(a) shows the studied multilayer heterostructure, which consists of perpendicularly magnetized (Co 0.4 nm /Pt 0.4 nm)$_2$ multilayers / 10 nm $Bi_{0.8}Sb_{0.15}$ topological insulator layer, capped by 1 nm MgO / 1 nm Pt (not shown). The multilayers were deposited on *c*-plane sapphire substrates by a combination of direct current (DC) and radio-frequency (RF) magnetron sputtering in a multi-cathode chamber. Note that the (Co/Pt)$_n$ (n=2-6) multilayers have been frequently used in MRAM production as a part of the synthetic antiferromagnetic reference layer with large PMA for pinning. Here, we chose the (Co 0.4 nm /Pt 0.4 nm)$_2$ multilayers (referred below as CoPt) to realize a thin ferromagnetic layer with a large PMA for evaluating the SOT performance of the BiSb layer. Characterization by X-ray diffraction and transmission electron microscopy shows that the BiSb layer is polycrystalline with the dominant (110) orientation (see Suppl. Info. Section 1). We also evaluated several sputtered BiSb single layers on sapphire substrates and observed the existence of metallic surface states as well as insulating bulk states with a band gap more than 170 meV for thin (< 20 nm) BiSb films.[24]

For electrical measurements, we fabricated 25 μm-wide Hall bars by optical lithography. An optical image of a Hall bar device with the experiment setup is shown in Fig. 1(b). From the parallel resistor model, we estimate that the conductivity of the BiSb layer is $1.5\times10^5$ $\Omega^{-1}$m$^{-1}$, which is much higher than that of sputtered Bi$_{1-x}$Se$_x$, and close to that of MBE grown BiSb on GaAs(111)A substrates. This demonstrates that is possible to grow highly conductive BiSb topological thin films on top of perpendicularly magnetized metallic layers by the sputtering technique. Thanks to the high conductivity of the BiSb layer, 50% of the applied current flows into the BiSb and contribute to the SOT magnetization switching.

Figure 1(c) shows the magnetic hysteresis curves of the as-grown sample measured by a superconducting quantum interference device (SQUID) under an in-plane and out-of-plane external magnetic field, respectively. The saturation magnetization $M_{CoPt}$, normalised by the CoPt layer thickness ($t_{CoPt}$ = 1.6 nm), is 613 emu·cm$^{-3}$. The uniaxial anisotropy field $H_u$ = 15 kOe is very large for the as-grown film. After Hall bar device fabrication by optical lithography undergoing several circles of thermal annealing, $H_u$ is reduced to 5.2 kOe (see Suppl. Info. Section 2), but is still much larger than that of NiFe or CoTb used in previous works. This CoPt layer yields a thermal stability factor for magnetization switching of Δ = 38, similar to that of the CoFeB(/MgO) free layer of perpendicular MTJs with diameters of 20-40 nm.[25,26] Figure 1(d) shows the anomalous Hall resistance ($R_H$) measured for a Hall bar under a sweeping out-of-plane field, which confirms PMA of the CoPt layer.

**Evaluation of the spin Hall angle by the second harmonic Hall measurements**

Next, we performed the second harmonic Hall measurements to evaluate the spin Hall angle.[27] An alternating current (AC) $J = J_0 \sin\omega t$ ($\omega = 259.68$ Hz) was applied to the Hall bar under a sweeping external field along the $x$ direction. We measured the 1st harmonic Hall resistance $R_H^\omega$, which indicates the anomalous Hall effect, and the 2nd harmonic Hall resistance $R_H^{2\omega}$, which is originated from the oscillation of the net magnetic moment under the spin-orbit effective magnetic fields.[28,29,30] Figure 2(a) and 2(b) show representative $R_H^\omega$ - $H_x$ and $R_H^{2\omega}$ - $H_x$ curves measured at $J^{BiSb} = 3.6 \times 10^5$ Acm$^{-2}$. In TI/FM systems, the spin Hall effect dominates the Rashba-Edelstein effect,[8,9,21] thus the second harmonic Hall resistance $R_H^{2\omega}$ can be expressed as[30]

$$R_H^{2\omega} = \frac{R_H}{2} \frac{H_{AD}}{H_x - H_u(H_x/|H_x|)} + R_{thermal} \frac{H_x}{|H_x|} \quad (1)$$

where $H_{AD}$ is the antidamping-like effective field and $R_{thermal}$ is the contribution from the anomalous Nernst (ANE) and spin Seebeck (SSE) effects. Fitting Eq. (1) to the high field data in the $R_H^{2\omega}$- $H_x$ curve yields $H_{AD}$ (red curves in Fig. 2(b)). Figure 2(c) shows $H_{AD}$ as a function of $J^{BiSb}$. From the $H_{AD}$ / $J^{BiSb}$ gradient, we can calculate $\theta_{SH} = \frac{2e}{\hbar} M_{CoPt} t_{CoPt} \frac{H_{AD}}{J^{BiSb}} = 12.3$, where $e$ is the electron charge, $\hbar$ is the reduced Plank constant. The obtained large $\theta_{SH} = 12.3$ and $\sigma = 1.5 \times 10^5$ Ω$^{-1}$m$^{-1}$ for sputtered BiSb thin film demonstrate the feasibility of BiSb for ultralow power SOT-MRAM.

**Ultrahigh efficient spin-orbit torque magnetization switching by DC and pulse currents**

Next, we demonstrate ultrahigh efficient and robust SOT magnetization switching in the CoPt/BiSb multilayers. Figure 3 shows the SOT magnetization switching by DC currents with an applied external field along the *x* direction. We achieved Hall resistance switching whose amplitude is consistent with that of the Hall resistance loop shown in Fig. 1(d), indicating full magnetization switching. The switching direction is reversed when the external magnetic field direction is reversed, which is consistent with the characteristic of SOT. Typical DC threshold switching current density $J_{th}^{BiSb}$ is $1.5 \times 10^6$ Acm$^{-2}$ at the bias field of 2.75 kOe. Note that thanks to the high electrical conductivity $\sigma = 1.5 \times 10^5$ $\Omega^{-1}$m$^{-1}$ of BiSb, the total current density including the shutting current in the CoPt is kept at $2.6 \times 10^6$ Acm$^{-2}$. Next, we performed SOT magnetization switching by pulse currents. Figure 4(a) and 4(b) show a representative SOT switching loops by 0.1 ms pulse currents at +1.83 kOe and -1.83 kOe, respectively. Figure 4(c) plots $J_{th}^{BiSb}$ at various pulse width $t_{pulse}$, and the theoretical fitting using the thermal activation model $J_{th}^{BiSb} = J_0^{BiSb} \times \left[1 - \frac{1}{\Delta} \ln\left(\frac{\tau_{pulse}}{\tau_0}\right)\right]$,[31] where $J_0^{BiSb}$ is the zero-kelvin threshold switching current density, $\Delta$ is the thermal stability factor, and $1/\tau_0 = 1$ GHz ($\tau_0 = 1$ ns) is the attempt switching frequency. The fitting yields $J_0^{BiSb} = 4.6 \times 10^6$ Acm$^{-2}$ and $\Delta = 38$. Since the obtained $\Delta$ is similar to that of the CoFeB(/MgO) free layer in perpendicular MTJ with size of 20~30 nm, our data can be used to estimate the performance of SOT-MRAM in such small sizes. For example, we estimate that the switching current density at $t_{pulse} = 10$ ns would be $J_{th}^{BiSb}$ (10 ns) = $4.3 \times 10^6$ Acm$^{-2}$, which is about 20 times smaller than that of heavy metals. Finally, we demonstrate robust SOT switching in the CoPt/BiSb junction. For this purpose, we applied a sequence of 75 pulses ($J_{th}^{BiSb} = 4.4 \times 10^6$ Acm$^{-2}$, $t_{pulse} = 0.1$ ms) as shown in the top panel of Fig. 4(c). The Hall resistance data recorded for a total of 150 pulses under ±1.83 kOe are shown in the bottom panel in Fig. 4(c). We observed a robust SOT switching with no change in the device characteristics, indicating that the BiSb topological insulator deposited by the sputtering technique has great potential for realistic SOT-MRAM.

**SOT performance benchmarking and conclusion**

Table 1 summarizes $\theta_{SH}$, $\sigma$, the spin Hall conductivity $\sigma_{SH} = (\hbar/2e)\sigma\theta_{SH}$, and the SOT normalized power consumption $P_n$ at room temperature of several heavy metals and TIs. Here, $\theta_{SH}$ of TIs are their best values reported in literature. For the calculation of the $P_n$, we assumed bilayers of spin Hall material (thickness $t$ = 6 nm for heavy metals and $t$ = 10 nm for TIs) and CoFeB (thickness $t_{FM}$ = 1.5 nm, conductivity $\sigma_{FM} = 6\times10^5$ $\Omega^{-1}$ m$^{-1}$). Considering the shunting current in the ferromagnetic layer, the SOT power consumption is proportional to $(\sigma t + \sigma_{FM}t_{FM})/(\sigma t\theta_{SH})^2$. One can see that not only $\theta_{SH}$ but also $\sigma$ affect the SOT power consumption, a fact usually overlooked in literature. For example, while the sputtered Bi$_x$Se$_{1-x}$ has a much larger spin Hall angle ($\theta_{SH}$ = 18.6) than that ($\theta_{SH}$ = 3.5) of MBE-grown Bi$_2$Se$_3$, their power consumption is nearly the same, because Bi$_x$Se$_{1-x}$ has poorer crystal quality than Bi$_2$Se$_3$ and thus very low conductivity. Meanwhile, the sputtered BiSb thin film in this work shows both high $\sigma = 1.5 \times 10^5$ $\Omega^{-1}$ m$^{-1}$ and large $\theta_{SH}$ = 12.3. Note that the obtained $\theta_{SH}$ is still smaller than the highest $\theta_{SH} \sim 52$ observed in the MBE-grown BiSb(012), because BiSb deposited on top of Pt is polycrystalline and does not have the optimized (012) orientation. We expect that even higher $\theta_{SH}$ can be obtained if we can control the crystal orientation of BiSb by inserting a seed layer that promotes the (012) orientation. Nevertheless, the obtained $\sigma_{SH} = 1.8 \times 10^6$ ($\hbar/2e$) $\Omega^{-1}$ m$^{-1}$ for the sputtered BiSb thin film in this work already outperforms other materials by one to two orders of magnitude. This high spin Hall performance of sputtered BiSb helps reduce the power consumption by nearly three orders of magnitude compared with that of W, which is the most used heavy metal for SOT-MRAM development. Our results demonstrate the feasibility of BiSb for not only ultralow power SOT-MRAM but also other SOT-based spintronic devices, such as race-track memories[32] and spin Hall oscillators.[33,34]

# Method

## Material growth

We deposited multilayers of (0.4 nm Co / 0.4 nm Pt)$_2$ / 10 nm Bi$_{0.8}$Sb$_{0.15}$ topological / 1 nm MgO / 1 nm Pt on *c*-plane sapphire substrates by DC (for Co, Pt, BiSb) and RF (for MgO) magnetron sputtering in a multi-cathode chamber. All layers are deposited by sputtering from their single targets using Ar plasma without breaking the vacuum at room temperature.

## Device fabrication

The samples were patterned into 90 μm-long × 25 μm-wide Hall bars by optical lithography and lift-off. A 45 nm-thick Pt were deposited as electrodes by DC magnetron sputtering, which reduces the effective length of the devices to 50 μm.

## SOT characterization

The samples were mounted inside a vacuumed cryostat equipped with an electromagnet. For the second harmonic measurements, a NF LI5650 lock-in amplifier was employed to detect the first and the second harmonic Hall voltages under sine wave excitation generated by a Keithley 6221 AC/DC current source. For the DC (pulse) current-induced SOT magnetization switching, a Keithley 2400 SourceMeter (6221 AC/DC current source) was used, and the Hall signal was measured by a Keithley 2182A NanoVoltmeter.

# Data availability

The data that support this study results are available from the corresponding author upon reasonable request.

# Acknowledgement

This work is supported by the CREST program of the Japan Science and Technology Agency (No. JPMJCR18T5). We sincerely thank N. Hatakeyama and H. Iida at the Material Analysis Division of Tokyo Institute of Technology for their help in TEM and XRD measurements.

**Table 1.** Spin Hall angle $\theta_{SH}$, electrical conductivity $\sigma$, spin Hall conductivity $\sigma_{SH}$, and SOT normalized power consumption $P_n$ of several heavy metals and topological insulators.

| SOT Materials | $|\theta_{SH}|$ | $\sigma$ ($\Omega^{-1}m^{-1}$) | $|\sigma_{SH}|$ [($\hbar/2e$) $\Omega^{-1}m^{-1}$] | $P_n$ |
|---|---|---|---|---|
| Ta | 0.15 | 5.3×10$^5$ | 8.0×10$^4$ | 1 |
| Pt | 0.08 | 4.2×10$^6$ | 3.4×10$^5$ | 3.6×10$^{-1}$ |
| W | 0.4 | 4.7×10$^5$ | 1.9×10$^5$ | 1.6×10$^{-1}$ |
| (Bi$_{0.07}$Sb$_{0.93}$)$_2$Te$_3$ (MBE) | 2.5 | 1.8×10$^4$ | 4.5×10$^4$ | 3.0×10$^{-1}$ |
| Bi$_2$Se$_3$ (MBE) | 3.5 | 5.7×10$^4$ | 2.0×10$^5$ | 2.1×10$^{-2}$ |
| Bi$_x$Se$_{1-x}$ (Sputtered) | 18.6 | 7.8×10$^3$ | 1.5×10$^5$ | 2.6×10$^{-2}$ |
| Bi$_{0.85}$Sb$_{0.15}$ (Sputtered) | 12.3 | 1.5×10$^5$ | 1.8×10$^6$ | 3.9×10$^{-4}$ |

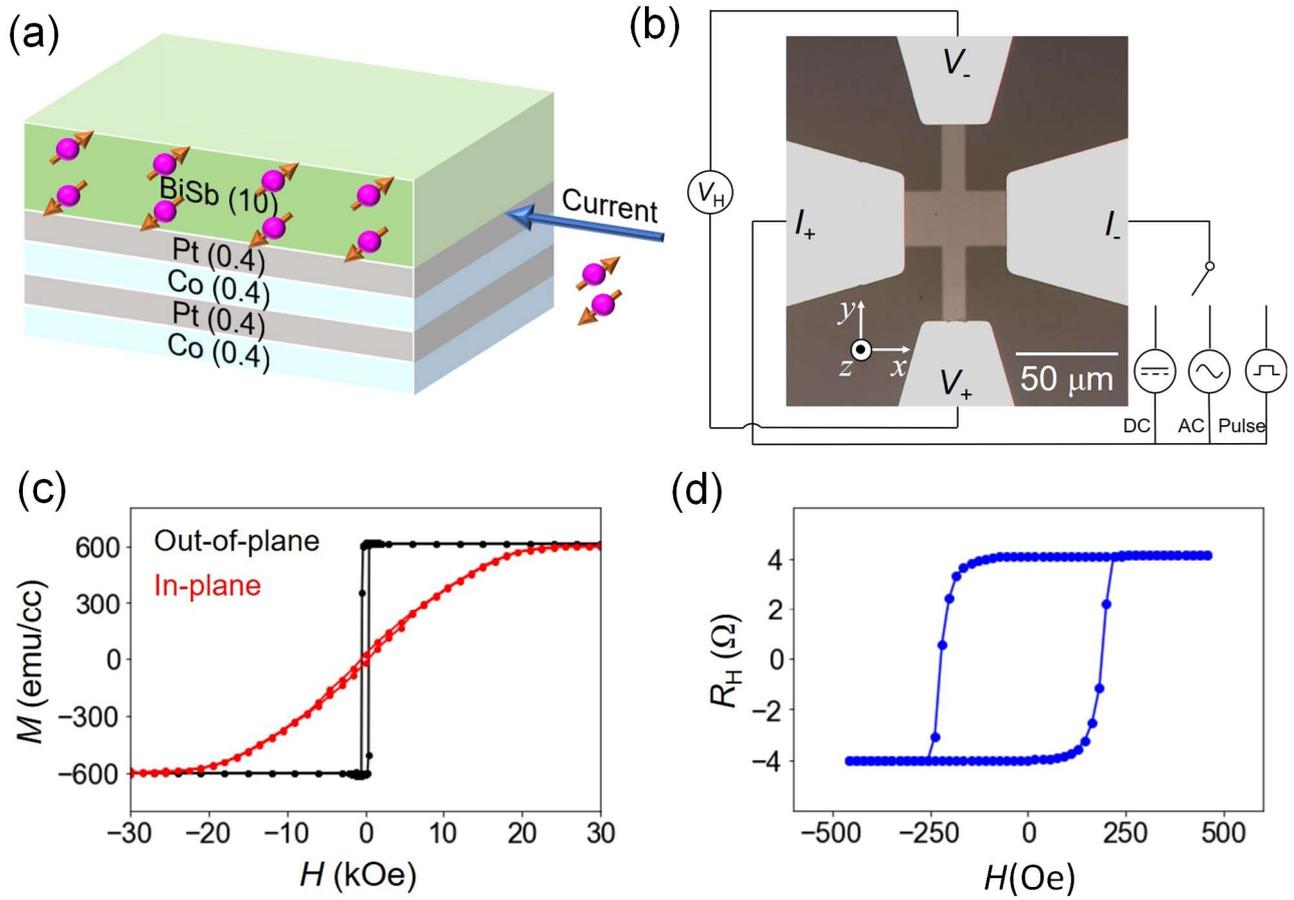

**Figure 1. Perpendicularly magnetized Co/Pt ferromagnetic multilayers – topological insulator BiSb heterostructure. (a)** Schematic structure of our multilayers. **(b)** Optical image of a Hall bar device and measurement configuration. **(c)** Magnetization curves of the Co/Pt multilayers. **(d)** Hall resistance of a Hall bar device measured with a perpendicular magnetic field.

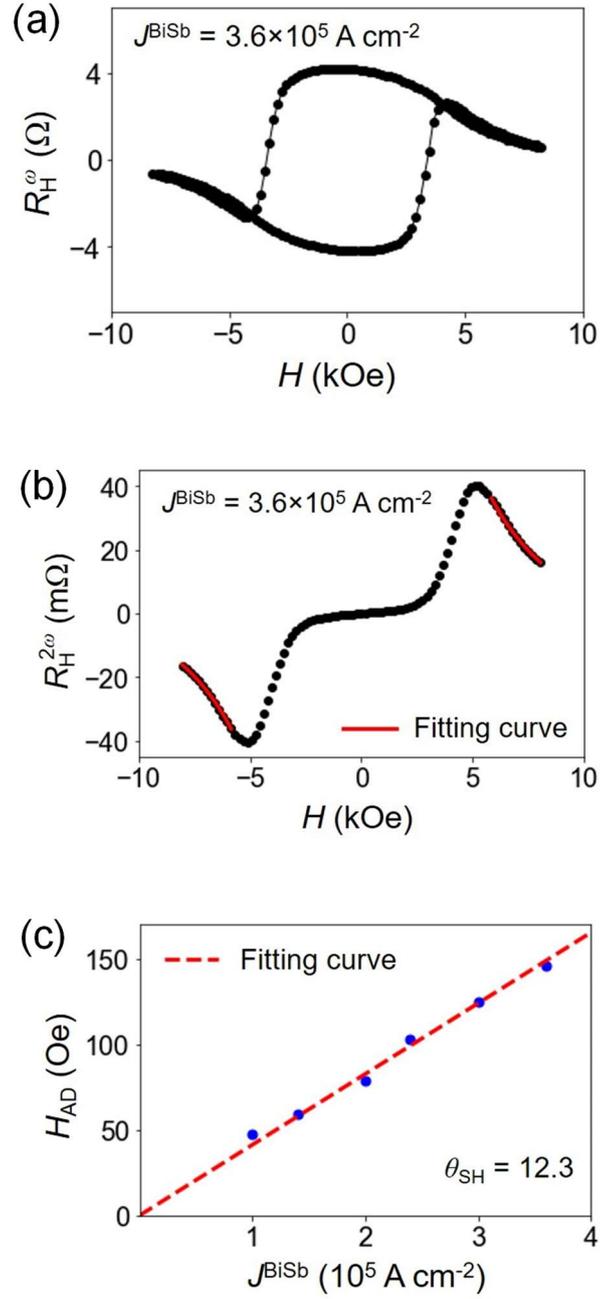

**Figure 2**. **Evaluation of the spin Hall angle by the second harmonic measurements.** **(a)(b)** 1st and 2nd harmonic Hall resistance as a function of the in-plane external magnetic field $H$, respectively. The red curves are the theoretical fitting using Eq. (1). **(c)** $H_{AD}$ as a function of $J^{BiSb}$.

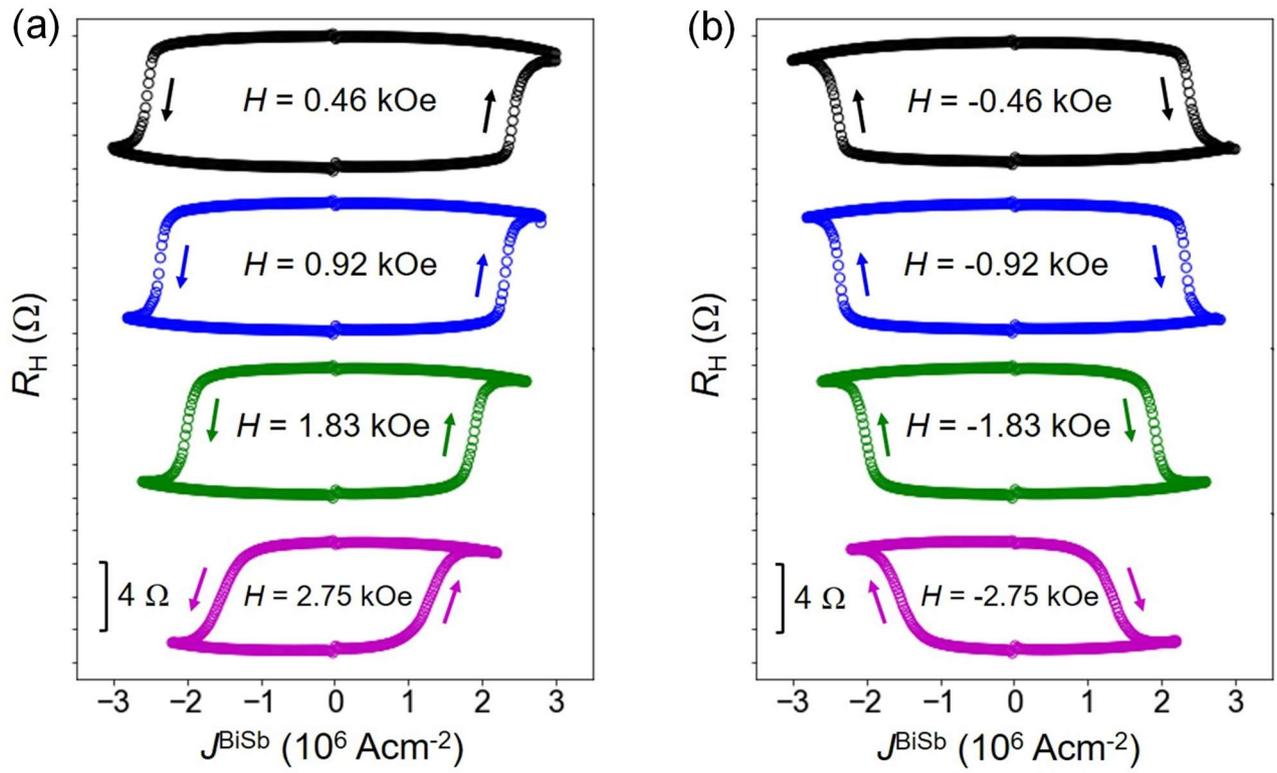

**Figure 3**. **SOT magnetization switching by DC currents.** Switching loops measured under an in-plane magnetic field applied along **(a)** +$x$ direction and **(b)** -$x$ direction.

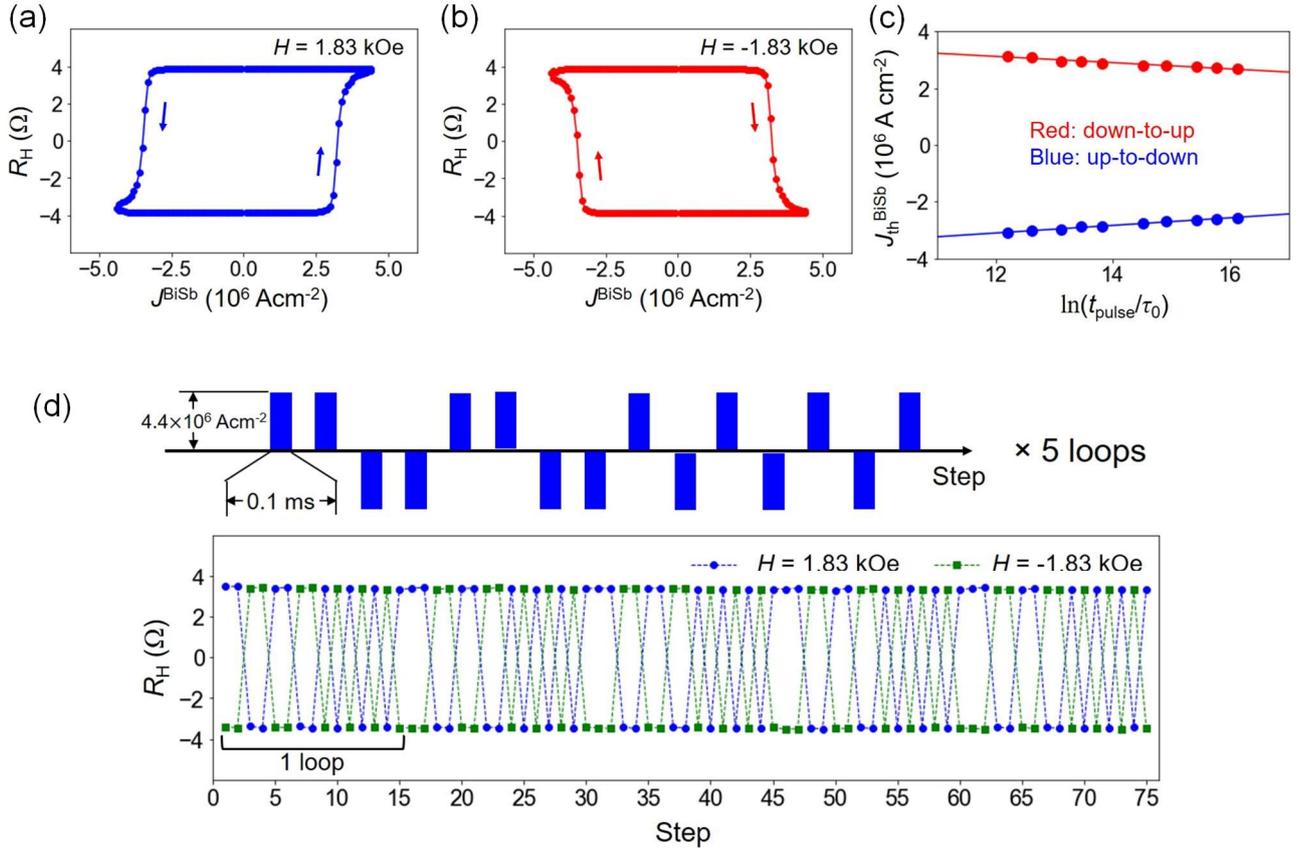

**Figure 4. SOT magnetization switching by pulse currents**. **(a)(b)** Switching loop by 0.1 ms pulse currents under an in-plane magnetic field of $H$ = +1.83 kOe and -1.83 kOe, respectively. **(c)** Threshold current density $J_{th}^{BiSb}$ as a function of $t_{pulse}$. **(d)** Robust SOT magnetization switching by 0.1 ms pulse current.

17